\documentclass[graybox]{svmult}

\usepackage{mathptmx}       
\usepackage{helvet}         
\usepackage{courier}        
\usepackage{type1cm}        
\usepackage{makeidx}         
\usepackage{graphicx}        
\usepackage{multicol}        
\usepackage{enumerate}
\usepackage[bottom]{footmisc}

\makeindex             

\begin{document}

\title*{Hipparcos Variable Star Detection and Classification Efficiency}
\author{Name of First Author and Name of Second Author}

\author{P. Dubath, I. Lecoeur-Ta\"ibi,
L. Rimoldini,
M. S\"uveges,
J. Blomme,
M. L\'opez,
 L. M. Sarro,
J. De Ridder,
 J. Cuypers,
L. Guy,
K. Nienartowicz,
A. Jan,
M. Beck,
 N. Mowlavi,
P. De Cat,
T. Lebzelter 
and L. Eyer}
\authorrunning{P. Dubath,
I. Lecoeur-Ta\"ibi,
L. Rimoldini et al.}


\institute{P. Dubath \at Observatoire astronomique de l'Universit\'e
  de Gen\`eve/ISDC,  ch. d'\'Ecogia 16, 1290 Versoix, Switzerland \email{Pierre.Dubath@unige.ch}}
%
%
\maketitle

\abstract{A complete periodic star extraction and classification
  scheme is set up and tested with the Hipparcos catalogue. The
  efficiency of each step is derived by comparing the results with
  prior knowledge coming from the catalogue or from the literature. A
  combination of two variability criteria is applied in the first step
  to select 17\,006 variability candidates from a complete sample of
  115\,152 stars.  Our candidate sample turns out to include 10\,406
  known variables (i.e., 90\% of the total of 11\,597) and 6600
  contaminating constant stars. A random forest classification is used
  in the second step to extract 1881 (82\%) of the known periodic
  objects while removing entirely constant stars from the sample and
  limiting the contamination of non-periodic variables to 152 stars
  (7.5\%).  The confusion introduced by these 152 non-periodic
  variables is evaluated in the third step using the results of the
  Hipparcos {\em periodic} star classification presented in a previous
  study (Dubath et al.~\cite{paper1}).}


\section{Introduction}
\label{sec:intro}

\begin{figure}[h]
\sidecaption
\includegraphics[width=\columnwidth]{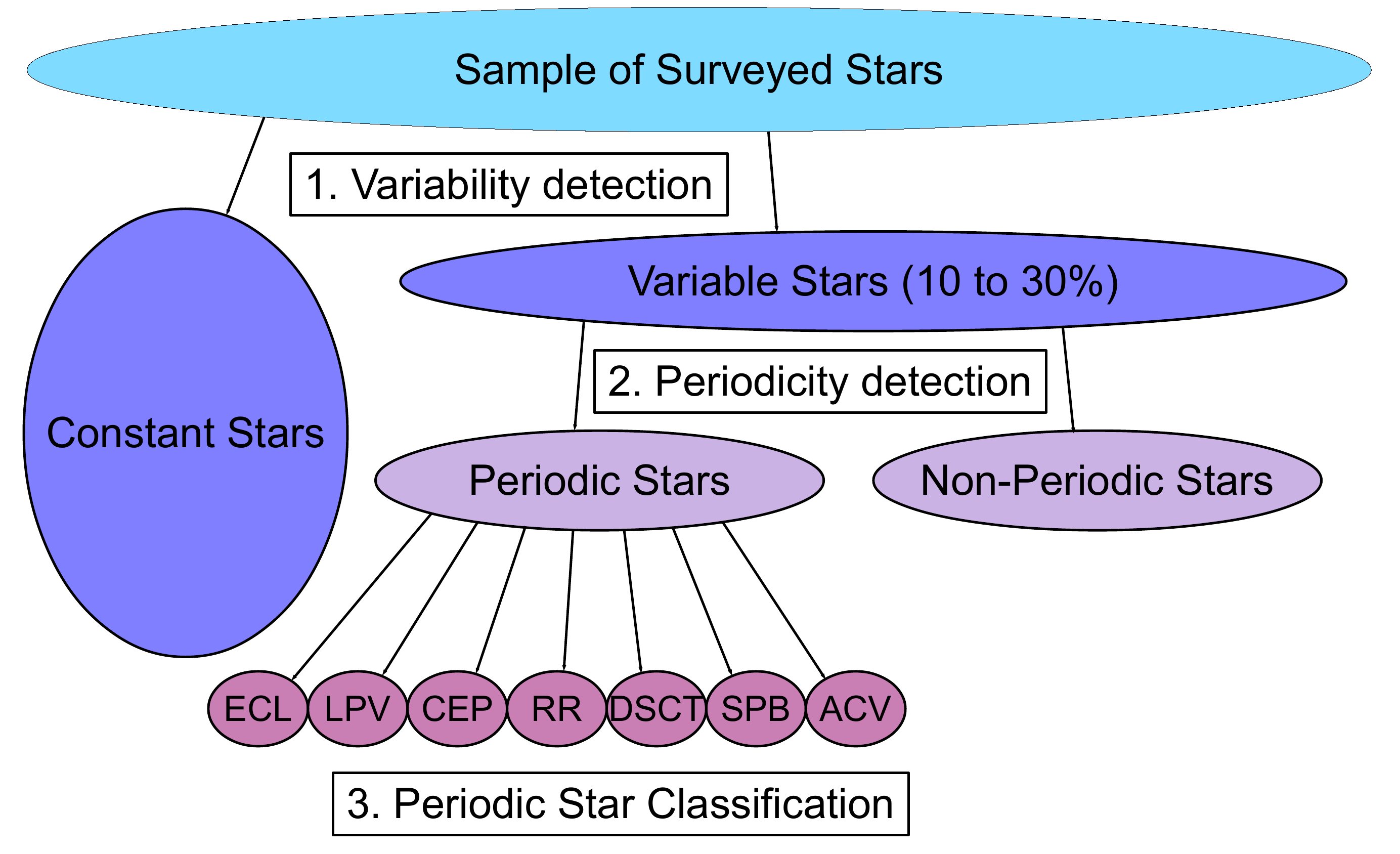}
 %
%
\caption{Illustration of the steps used in this study to identify and
  classify variable sources}
\label{fig:1}       
\end{figure}

Current and forthcoming photometric surveys are monitoring very large
numbers of astronomical targets providing a fantastic ocean for fishing
interesting variable objects. However, because of the large numbers
involved, their extraction requires the use of fully automated and
efficient data mining techniques. In this contribution, we use the
Hipparcos data set to investigate the performance of a complete and
automated scheme for the identification and the classification of
periodic variables. As shown in Fig. \ref{fig:1}, we study a three
step process. In the first step variable candidates are separated from
the objects most likely to be constant. This saves significant
processing time as period search is performed only in the subset of
variable candidates. The validity of the detected periods is
established in the second step, which separates truly periodic from
non-periodic objects. The third step is the classification of periodic
variables into a list of types (only a sub-set of them are shown in
Fig. \ref{fig:1}). This step is presented in details in Dubath et
al.~\cite{paper1}.  To avoid unnecessary repetition, this first paper
is referred to for a full description of the classification attribute
calculation and of the details of the random forest methodology.


This three step organisation represents a particular
option. Alternatives are also being considered, but they are outside
the scope of this contribution as is the classification of
non-periodic variables which is the subject of another study
(Rimoldini et al., in preparation).


\section{Variability Detection}
\label{sec:var_det}

In order to select variable star candidates, a number of variability
criteria are computed from the Hipparcos light curves\footnote{Only
  data point with quality flags 0 and 1 have been used in the light
  curves and stars with light-curves with less than 5 good data points
  are discarded.}. These criteria are all, in one way or another,
characterizing an excess of scatter compared to the one expected from
random noise. Some of them rely on the noise estimations while others
do not. P-values are computed for each of the tests. The star is
accepted as a variable candidate if the p-value is smaller than a
specified threshold.

%
\begin{figure}[h]
\sidecaption
\includegraphics[width=\columnwidth]{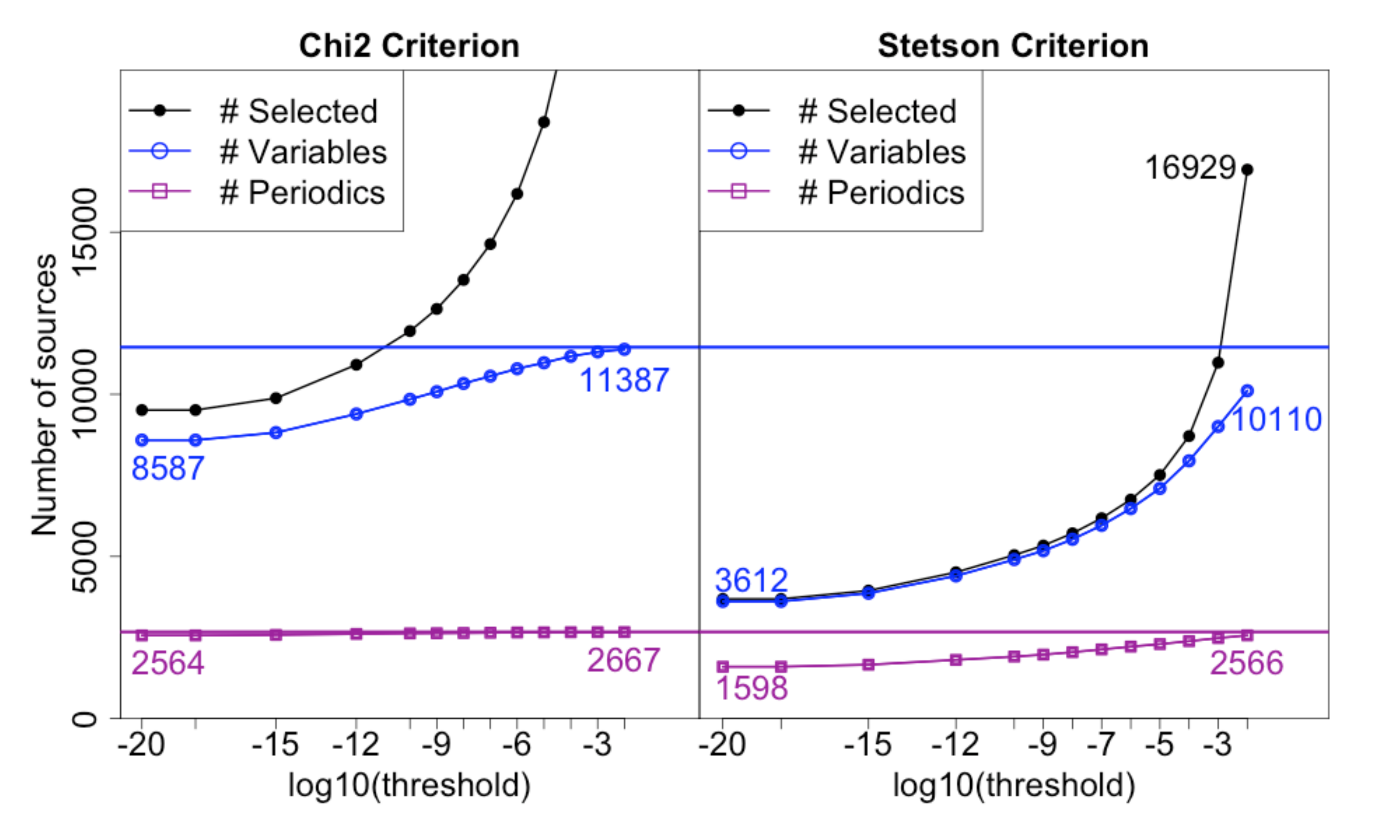}
 %
%
\caption{Number of stars selected using the chi-square criterion (left)
  and the Stetson~\cite{stetson} one (right) as a function of the p-value
  threshold. The total numbers of selected stars are displayed in
  black, while the fraction of periodic and variable stars are shown
  in magenta and blue, respectively. The numbers of variable stars drawn
  in blue include the contribution of periodic and non-periodic
  objects. The difference between the blue and the black curves
  indicates the amount of contamination from non-variable stars. The
  two horizontal lines indicate the total number of periodic stars
  (2672 in magenta) and of variable stars (11\,453 in blue). The complete
  sample includes 115\,152 stars.}
\label{fig:2}       
\end{figure}

Figure \ref{fig:2} shows the number of selected sources as a function
of the p-value threshold obtained from a chi-square criterion in the
left panel and from an alternative criterion proposed by
Stetson~\cite{stetson} in the right one.




As expected, the number of selected stars increases with larger
p-values thresholds in both panels. The optimum threshold maximizes
the numbers of selected true variables while limiting the
contamination by false positives (i.e., by constant stars). The
chi-square criterion is efficient at finding variables, but it also
includes a large number of false positives, even when the threshold is
extremely small. This suggests that Hipparcos photometric errors may
be slightly underestimated. The Stetson criterion is quite efficient
for periodic variables and it limits better the number of false
positive detections, but it misses more non-periodic variable stars.

%
\begin{figure}[h]
\sidecaption
\includegraphics[width=\columnwidth]{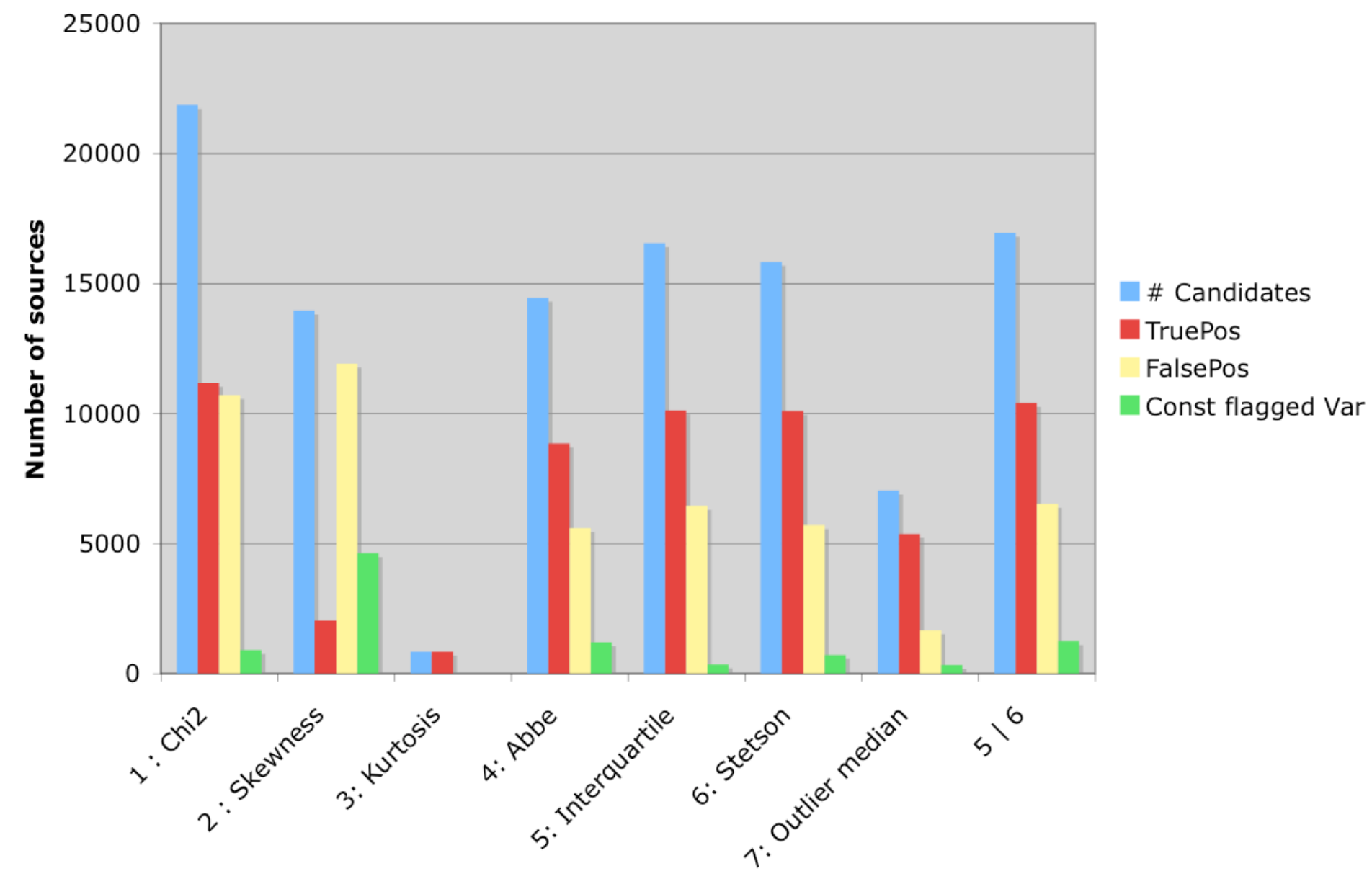}
 %
%
\caption{Comparison of the numbers of stars selected using different
  variability criteria and a particular near-optimum p-value
  threshold. The total numbers of variable candidates appear in blue,
  red bars show the number of candidates flagged in the Hipparcos
  catalogue as variables (i.e., the true positive detections), yellow
  bars indicates the number of false positives. The fraction of false
  positives flagged as ``constant'' in the Hipparcos catalogue are
  shown in green. }
\label{fig:3}       
\end{figure}

Figure~\ref{fig:3} shows a comparison of the numbers of stars selected
using different variability criteria tested with a particular
near-optimum p-value threshold. The variability criteria tested
include (1) the chi-square criterion, (2) the skewness and (3) the
kurtosis of the magnitude distributions, the (4) Abbe criterion (e.g.,
see Strunov~\cite{strunov}), (5) the interquartile range, (6) the Stetson
criterion, (7) the outlier median criterion and (8) the union of the
Stetson and interquartile criteria.

This figure shows again that the chi-square is the most efficient
criterion at identifying variable stars, but that it also includes the
largest contribution of false positive detections.  A final sample of
17\,006 variable candidates (i.e, 14.8\% of the total) is formed by
merging the Stetson and the interquartile selections obtained with
p-value thresholds of 10$^{-2}$ and 10$^{-3}$, respectively. This
sample is used in the subsequent steps of this study.

\section{Periodicity Detection}

Fig.~\ref{fig:1} indicates that periodicity detection is the second
step. This figure might however be misleading as it assumes that the
first step is perfect. In reality, the second step starts with a
sample of variable star {\em candidates}, which includes a number of
constant objects. With the knowledge of the Hipparcos
catalogue\footnote{http://www.rssd.esa.int/index.php?project=HIPPARCOS\&page=Overview},
we know quite precisely what mixture of stars is included in our
selection. Out of the 17\,006 candidate sample, (1) 2657 stars are
flagged as periodic in the Hipparcos catalogue (flag H52), (2) 6954 as
unsolved\footnote{Stars flagged as ``Unsolved'' have Hipparcos light
  curves from which it was not possible to derive significant evidence
  for a period. They may include periodic stars with light-curves of
  insufficient quality or truly non-periodic sources. }, (3) 794
micro-variable, (4) 762 as constant and (5) 4360 are not flagged,
because they were not considered variable nor constant with any degree
of confidence\footnote{662 stars flagged as ``R'' (for revised color
  index) and 816 stars flagged as ``D'' (duplicity-induced
  variability) in the Hipparcos catalogue are not included in our
  training set (see page 121 of the Hipparcos catalogue).}. These stars are
used to train and test the performance of a random forest supervised
classifier for identifying periodic variables.

Using the procedures and criteria described in section 3 of Dubath et
al.~\cite{paper1}, a good period is obtained for 2323 of the 3022
stars with a known period included in our 17\,006 star sample (i.e., a
good period recovery rate of 77\%). There are 357 stars flagged as
periodic with wrong period values. Those are eliminated from our
training set as well as the 20 ``unsolved'' stars for which a good
period value is found.

\begin{figure}[h]
\sidecaption
\includegraphics[width=\columnwidth]{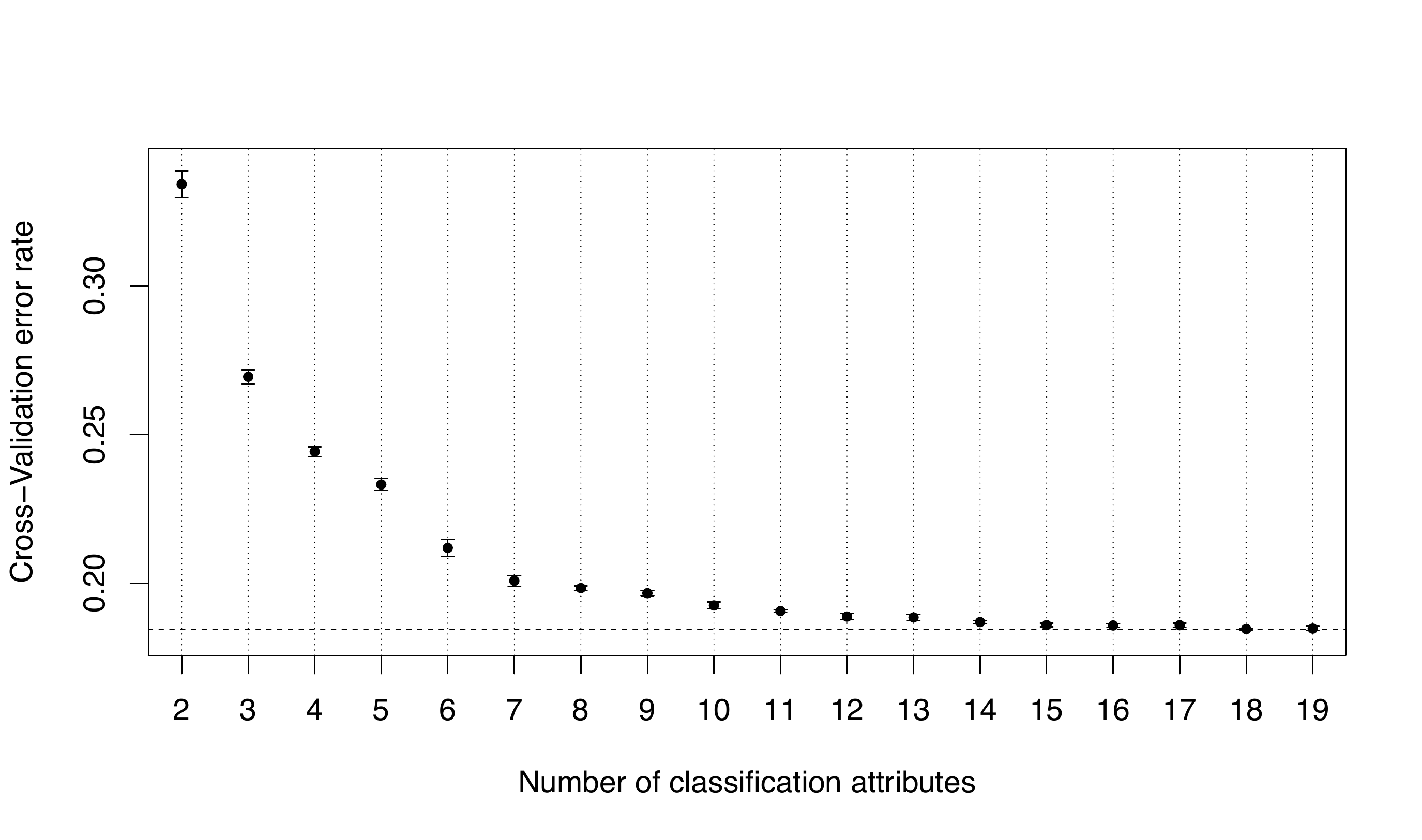}
 %
%
\caption{Evolution of the Cross-Validation (CV) error rate as more and
  more attributes are added into the classification process. In the
  different CV experiments (hundred for each attribute number), the exact
  list of attributes used to estimate the error rate is not always
  exactly the same (see Fig. \ref{fig:5}).}
\label{fig:4}       
\end{figure}

A large number of attributes are computed and the procedure presented
in section 4 of Dubath et al.~\cite{paper1} is followed to rank and
select the most important attributes. Figure~\ref{fig:4} displays the
results of a series of ten experiments of 10-fold Cross-Validation
(CV), i.e., 100 experiments for each attribute number.
 
%

Figure~\ref{fig:4} shows that the three most important attributes
already drive the mean error down to 27\%, which reduces to 20\% with
7 attributes. The mean error continues to decrease slowly until it
reaches a plateau of 18.5\%. Using more than about 15 attributes does
not lead to further significant improvements.

\begin{figure}[h]
\sidecaption
\includegraphics[width=0.6\columnwidth]{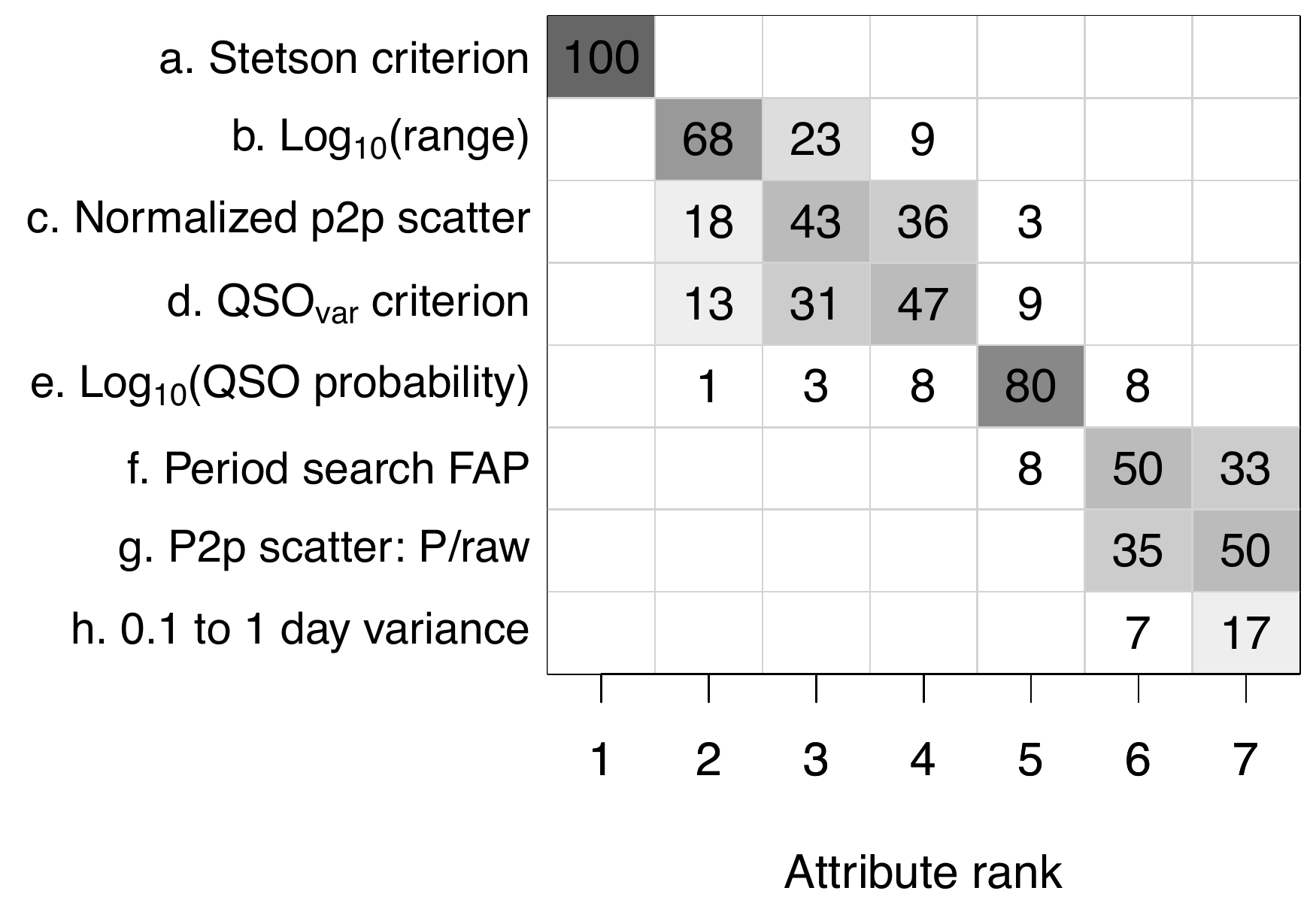}
 %
%
\caption{This figure shows the ranking of the most important
  attributes in the ten series of 10-fold cross-validation
  experiments.}
\label{fig:5}       
\end{figure}

Figure~\ref{fig:5} displays the ranking of the most important
attributes in the CV experiments. Fig.~\ref{fig:4} and \ref{fig:5}
should be read together. While Fig.~\ref{fig:4} shows that experiments
done with 3 attributes result in a mean error rate of 27\%,
Fig.~\ref{fig:5} indicates that most of the time the 3 most important
attributes are those labelled (a), (b), and (c). Numbers in this
figure indicate the number of time that the attribute has a
particular rank in the series of 100 experiments.

Below we provide a short description of the most important
attributes displayed in Fig.~\ref{fig:5}.

\begin{enumerate}[a]

\item Stetson criterion -- Stetson variability index~\cite{stetson}
  pairing successive measurements if separated by less than 0.05
  days. This time interval is optimized to be long enough to make many
  pairs while remaining much shorter than typical period values.

\item Log$_{10}$(range) -- decadic log of the range of the raw time series magnitudes.

\item Normalized p2p scatter -- Point-to-point scatter computed on the
  folded time-series normalized by the mean of the square of the
  measurement errors.

\item QSO$_{\rm var}$ criterion -- Reduced $\chi^2$ of the source
   variability with respect to a parametrized quasar variance model
   (denoted by $\chi^2_{QSO}/\nu$ in Butler \& Bloom~\cite{butler+}).

\item Log$_{10}$(QSO probability) -- Log$_{10}$ of a quantity defined
  by Eq.~8 in Butler \& Bloom~\cite{butler+}.



 \item Period search FAP -- False-alarm probability associated with the maximum peak in the Lomb-Scargle periodogram.

 \item P2p scatter: P/raw -- Point-to-point scatter from folded time-series normalized by the same quantity computed on raw time-series.

\item Variance within 0.1 to 1 day intervals -- The average of
  absolute magnitude variations on time scales from 0.1 to 1 days.

\end{enumerate}

\begin{figure}[h]
\sidecaption
\includegraphics[width=0.6\columnwidth]{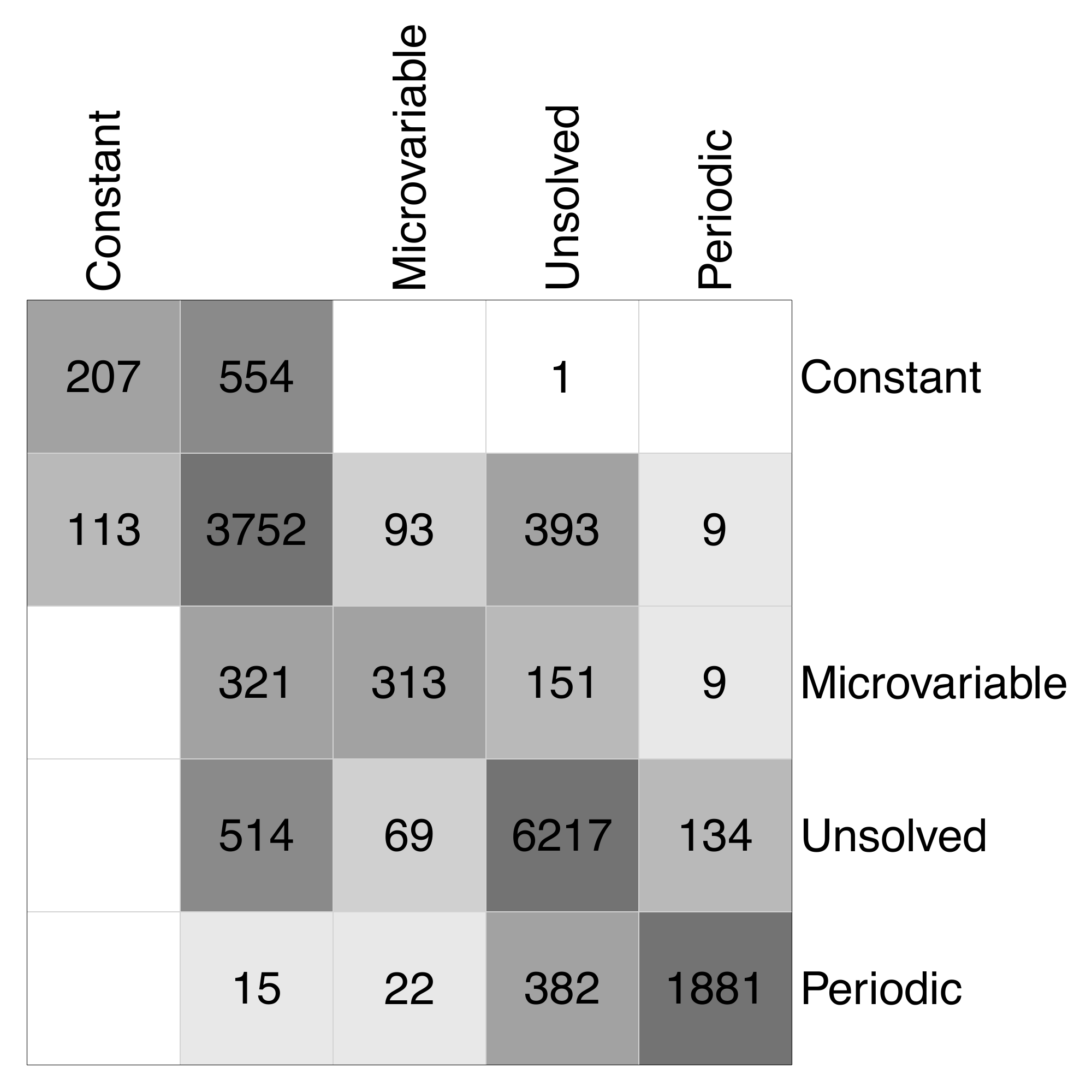}
 %
%
\caption{Confusion matrix obtained with the out-of-bag samples in a
  2000-trees random forest classification.}
\label{fig:6}       
\end{figure}

Figure~\ref{fig:6} shows the confusion matrix obtained from the
out-of-bag samples of a 2000-trees random forest classification. Out
of the 2300 periodic stars with good periods, 1881 (82\%) are
correctly identified while 419 (18\%) are missed, mostly appearing in
the ``unsolved'' category. Remarkably, only 152 stars (134 unsolved, 9
micro-variables and 9 stars without flags) are wrongly classified as
``periodic'', resulting in a total contamination of the periodic type
of 7.5\%. There is also no confusion between ``constant `` and
``periodic'' types.

\section{Impact on periodic star classification}

\begin{figure}[h]
\sidecaption
\includegraphics[width=\columnwidth]{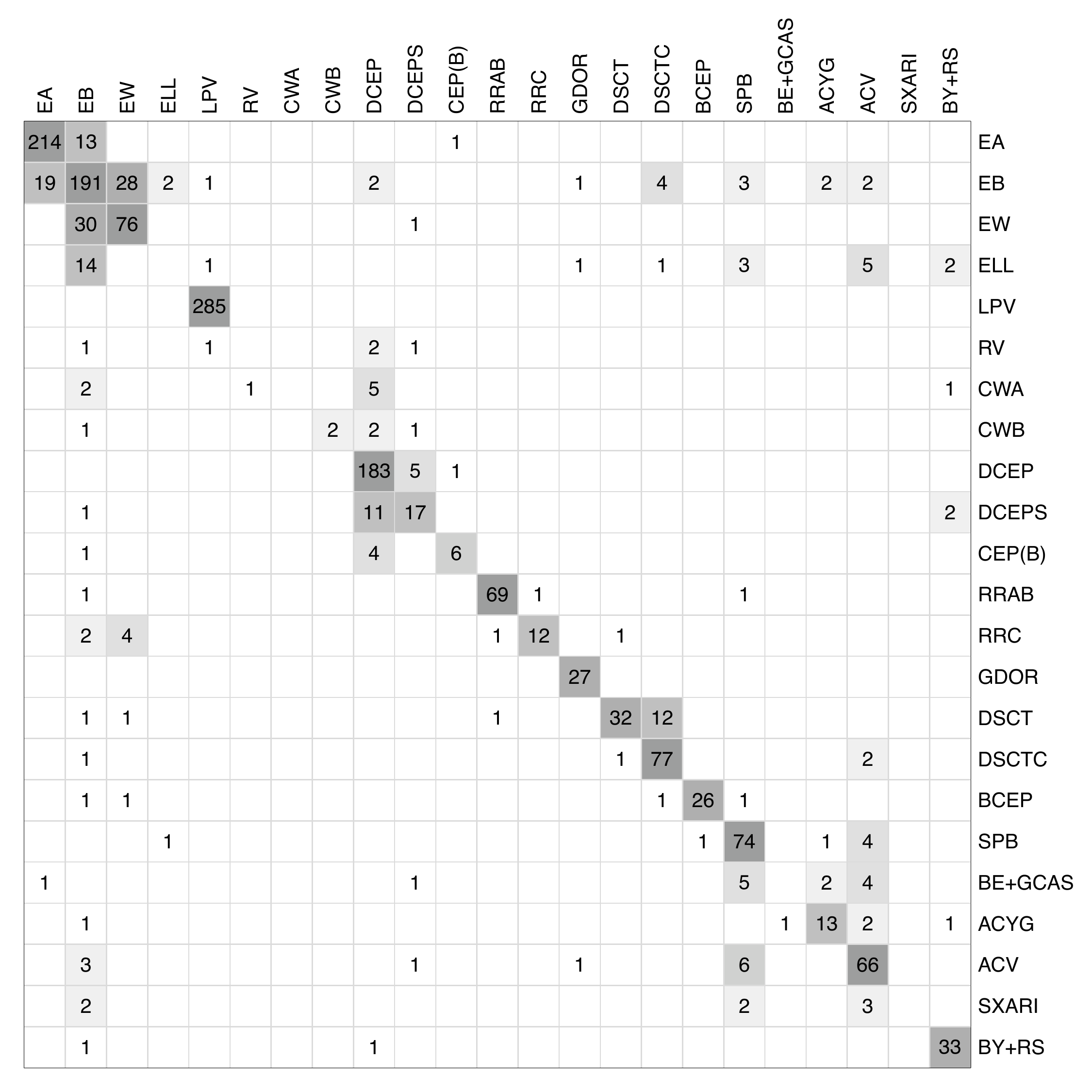}
 %
%
\caption{Confusion matrix obtained by Dubath et al.~\cite{paper1} for
  the Hipparcos periodic variable stars.}
\label{fig:7}       
\end{figure}

The classification of the Hipparcos periodic variable stars is the
subject of a previous study (Dubath et al.~\cite{paper1}). The
confusion matrix obtained in that study is displayed in
Fig.~\ref{fig:7}. This figure represents however an optimistic picture
as this sample only contains the best known stars, for which we have
relatively clean light-curves. It is very difficult to evaluate
accurately the extent of the expected degradation when using this model
to classify other stars. Some indications can however be derived in
two different ways.

First, the classification model derived from the training set can be
applied to the sample of Hipparcos stars with uncertain types from the
literature. The results of this process is shown in Fig.~10 and 11 by
Dubath et al.~\cite{paper1}, where a relatively mild confusion is
observed and evaluated.

Second, the current study shows that any sample of periodic stars is
expected to be contaminated by non-periodic stars because of the
imperfection of the two preliminary steps, namely variability and
periodicity detections. This contamination is evaluated to about 7.5\%
in last section (see Fig.~\ref{fig:6}). The 152 stars wrongly identified
as periodic can be classified using a periodic classification model to
evaluate more precisely the contamination in terms of periodic types.

A 10-fold cross-validation experiment is carried out to extract the
variables wrongly classified as periodic: 150 stars, including 133
"Unsolved", 8 "Micro-variable" and 9 with no flag. These numbers
slightly differ from the corresponding ones in Fig.~\ref{fig:6} due to
the randomness involved in random forest classification. The
classification model from Dubath et al.~\cite{paper1} is then used to
predict periodic types for these stars.

The predicted types for the 150 stars turn out to include 125
Long-Period Variables (LPVs), 9 RR Lyrae of type AB, 6 Delta Scuti and
eclipsing binaries of type EA (5) and EB (5). The LPV classification
prediction for 125 stars could easily be understood if they had large
amplitude, red color and long (most probably spurious) periods as
expected for such kind of stars. However, this is not supported by the
data. The understanding of the true nature of these stars
requires further investigation.




%
%
%


\end{document}